# Diffusion-Oscillatory Dynamics in Liquid Water on Data of Dielectric Spectroscopy

A. A. Volkov[1], V. G. Artemov[1(a)], A. A. Volkov[2], and N. N. Sysoev[2]

[1] *A.M. Prokhorov General Physics Institute, Russian Academy of Sciences - 119991, Moscow, Russia*
[2] *Physical Department, M.V. Lomonosov Moscow State University - 119991, Moscow, Russia*

(Dated: June 20, 2016)

When analyzing the broadband absorption spectrum of liquid water ($10^{10}$ - $10^{13}$ Hz), we find its relaxation-resonance features to be an indication of Frenkel's translation-oscillation motion of particles, which is fundamentally inherent to liquids. We have developed a model of water structure, of which the dynamics is due to diffusion of particles, neutral $H_2O$ molecules and $H_3O^+$ and $OH^-$ ions - with their periodic localizations and mutual transformations. This model establishes for the first time a link between the dc conductivity, the Debye and the high frequency sub-Debye relaxations and the infrared absorption peak at 180 cm$^{-1}$. The model reveals the characteristic times of the relaxations, 50 ps and 3 ps, as the lifetimes of water molecules and water ions, respectively. The model sheds light on the anomalous mobility of a proton and casts doubt on the long lifetime of a water molecule, 10 hours, commonly associated with autoionization.

PACS numbers: 77.22.-d, 66.10.Ed

## I. INTRODUCTION

The intermediate position of a liquid state between gases and solids naturally implies that atoms and molecules of a liquid participate simultaneously in the oscillatory and translational motions. The related physical model was offered by Frenkel [1] in the 1930s, and 30 years later was successfully used for the interpretation of data of neutron scattering in liquid water [2, 3]. It was assumed that each molecule of water participates in Brownian diffusion, being for a short time localized to oscillate in the enclosure (cage) of its neighbors.

It would seem that the combined translation-oscillatory motion of molecules in a liquid, owing to its fundamental generality, ought to have become a universal touchstone for developing dynamic water models, but this did not happen. Conventional knowledge considers water as an assemblage of tetrahedral gaseous $H_2O$ molecules, tightly held together by hydrogen bonds [4-6]. It is assumed that the fast breaking and forming of hydrogen bonds is what permits water molecules to move translationally and rotate correlatively to produce a high-value dielectric constant.

Many sophisticated models have been proposed to find a certain mechanism which is behind translation-rotational dynamics in liquid water [6-9]. Surprisingly, the solution of the problem went on for decades, and the uniform microscopic picture has still not been achieved [10]. The problem of the local water structure still challenges researches [11].

Our suggestion is to seek an adequate description of the water dynamics via consideration of Frenkel's translation-*oscillatory* motion rather than of common translation-*rotational* motion. We proceed from the famous Debye relaxation in water [4, 12, 13] being an ideological basis of the rotational motion does not exclude alternative interpretations [14]. For the first time we considered the possible alternative in the form of a translational movement of a charged particle in the parabolic potential [15].

In this study we develop a simple physical model of translational-oscillatory motion of particles ($H_2O$ molecules and $H_3O^+$ and $OH^-$ ions) which fully meets the dispersion features of the absorption spectrum of liquid water across a wide range of frequencies $10^3$ - $10^{13}$ Hz including the domain of the Debye relaxation.

## II. EXPERIMENTAL DATA

In Fig. 1, the frequency panorama of a dielectric response of water is presented. It is constructed according to the studies [16-19]. The top panel shows the spectrum of the real part of dielectric function $\varepsilon(\omega)$ and the lower panel presents the spectrum of dynamic conductivity $\sigma(\omega) = \omega\varepsilon_0\varepsilon''(\omega)$, where $\varepsilon''(\omega)$ is the imaginary part of dielectric permittivity, $\omega$ is the circular frequency, $\varepsilon_0$ is the dielectric permittivity of a vacuum. The absorption peak at 5.3 THz

(a)E-mail: vartemov@bk.ru

(marked by an arrow) divides the panorama into two parts. On the left, near ~ $2\cdot10^{10}$ Hz, the $\varepsilon(\omega)$ and $\sigma(\omega)$ graphs as two smooth knees represent the Debye relaxation [12, 13]. To the right of the 5 THz peak the $\sigma(\omega)$ spectrum is split by oscillations and looks similar to the infrared spectrum of an ionic crystal [20].

and $R_2$ and two lorentsians $L_1$ and $L_2$ [19]. In terms of conductivity the spectrum $\sigma(\omega)$ looks as:

$$\sigma(\omega) = \sigma_0 + \sum_i \frac{\sigma_i (\nu/\nu_i)^2}{1+(\nu/\nu_i)^2} + \sum_j \frac{\sigma_j \nu^2 (\nu_j \gamma_j)^2}{(\nu_j^2 - \nu^2)^2 + (\nu \gamma_j)^2}, \quad (1)$$

where i = 1, 2, j = 1, 2 and $\sigma_0$ is the dc conductivity $\sigma_{dc}$.

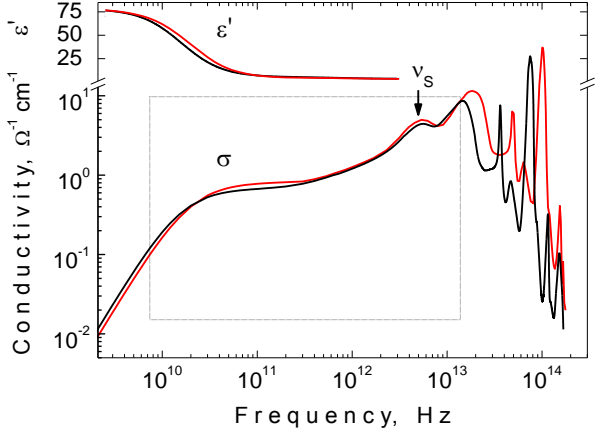

FIG. 1: (Color online) Spectra of permittivity, $\varepsilon(\omega)$, and dynamical conductivity, $\sigma(\nu)$, of liquid water at room temperature on data taken from refs. [16-19]; the red and black lines relate to light ($H_2O$) and heavy ($D_2O$) water, respectively. The arrow indicates the distinctive 5.3 THz peak (see text). The dashed frame shows a fragment presented in extended scale in Fig 2.

Fig. 1 shows the spectra of light and heavy water in comparison. They demonstrate an isotopic effect, i.e., the systematic shift of frequency peaks which, as shown in the diagram, has no effect on the 5 THz peak. Spectral shape is interpreted as being due to the two modes of molecular motions – intra- and inter-molecular [7, 20]. It is believed that the high-frequency resonances are due to the motion of lighter protons while the heavy oxygen atoms are frozen.

Located in the array of the intramolecular oscillations, the 5 THz peak is nevertheless associated with translational motion of molecules [7]. Besides tolerance to the isotope-effect it is indifferent to the temperature [21] and presence of electrolyte ions in water [22]. The peak is active in both infrared spectra and Raman spectra [20, 23].

The fragment of the $\sigma(\omega)$ spectrum, which contains the 5 THz resonance, is marked by shading in Fig. 1 and shown in Fig. 2 in a magnified view. It has been shown in many studies that this part of the spectrum is comprehensively described with an additive sum of two relaxators $R_1$

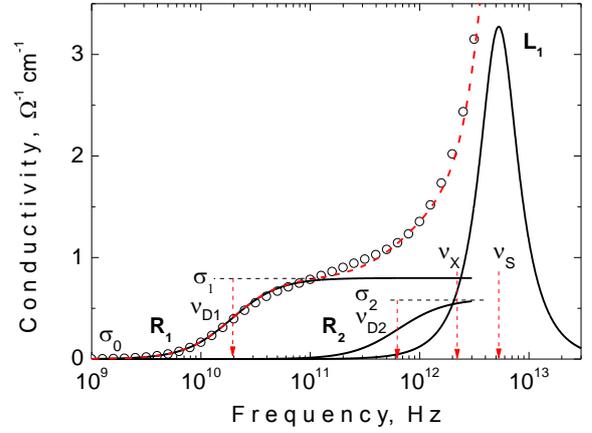

FIG. 2: (Color online) Fragment of the conductivity spectrum $\sigma(\nu)$ shown in Fig. 2 with a dashed frame – circles-dashed red line. The solid black lines represent the separate spectral contributions of the Debye relaxation, $R_1$, the sub-Debye relaxation, $R_2$, and the 5.3 THz peak, L. Other notations are explained in text.

The contours of $R_1$ and $R_2$ and $L_1$ are shown in Fig. 2. They are calculated for definiteness according to the study [19] with the parameters specified in Table 1.

TABLE 1: The parameters of the conductivity spectrum $\sigma(\omega)$ shown in Fig. 2 as input parameters of the model shown in Fig. 3. In addition to table, $\nu_X = 2.1\cdot10^{12}$ Hz and $\sigma_0 \equiv \sigma_{dc} = 5.5\cdot10^{-6}$ $\Omega^{-1}m^{-1}$ [24].

| Spectral component | $R_1$ | $R_2$ | $L_1$ |
|---|---|---|---|
| Subscript for ν | D1 | D2 | S |
| σ, $\Omega^{-1}m^{-1}$ | 80 | 140 | 320 |
| ν, $10^{12}$ Hz | 0.02 | 0.64 | 5.3 |
| γ, $10^{12}$ Hz | - | - | 5.4 |

As is seen, the relaxations $R_1$ and $R_2$ meet the 5 THz absorption peak at frequency $\nu_X$ = 2.1 THz. In the high frequency limit, to the right, the relaxations come to plateaus $\sigma_1$ = 80 and $\sigma_2$ = 140 $Ohm^{-1}m^{-1}$ while the relaxations stream from the 5 THz peak to the left as two absorption bands. The first band, $R_1$, is the above mentioned Debye relaxation, while the second band, $R_2$, is its satellite [12, 19]. Molecular nature of both relaxations is vague and still debatable [7, 25].

Below we construct a molecular structure (an instantaneous I-structure in terms of [4]) the dynamics of which would naturally meet the shape of the spectral $R_1$, $R_2$, $L_1$ bands.

## III. MODEL

The experimental absorption spectrum in Fig. 2 is reconstructed for convenience in Fig. 3 in the stylized form. As in the study [2, 3] we consider an assemblage of $H_2O$ molecules each of them making an oscillatory motion and also moving as free diffusing particle. In addition, we introduce consideration of $H_3O^+$ и $OH^-$ ions (protonated and deprotonated $H_2O$ molecules) to make diffusion observable in the conductivity. We believe ions to be indicators of general particle motion giving the resonant and relaxation response in the conductivity spectrum.

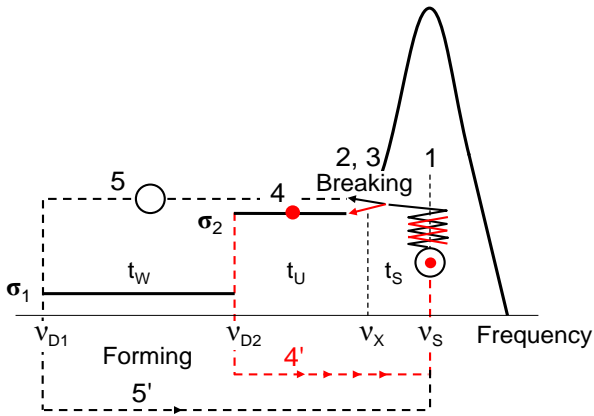

FIG. 3: (Color online) The model conductivity spectrum of liquid water $\sigma(\nu)$ - thick solid line. Open circle is an $H_2O$ molecule, red point is an excess proton, point in circle is an $H_3O^+$ ion. Dashed lines show the life cycles of a water molecule (black) and a charge (red). Other notations are explained in text. Digits are that used in Fig. 4.

The fundamentally important feature of our model is the possibility for a proton to transfer from one oxygen to another during the thermal ion-molecular collision. Thus, the model is similar developed in works [2, 3], but differs in that includes the proton exchange between particles.

The figure (1) in Figs. 3 and 4 indicates the oscillation motion of an $H_3O^+$ ion in a hydrate cage with a thermal frequency of $\nu_S$. We will call such ions surrounded with a hydrate cage "dressed", their concentration $N_S$. The hydration cage is a center-symmetric configuration of polarizationaly oriented $H_2O$ molecules (polarization in Fig. 4 is not shown for the sake of simplicity).

Let us denote the concentrations of molecules and ions as $N_W$ и $N_I$, respectively. A snapshot of the structure is shown in Fig. 4. The digits in Figs. 3 and 4 coincide. They represent the elementary processes.

An excess proton or a hole supplies a neutral $H_2O$ molecule with a charge which transforms it into an ion and makes it active in the absorption spectrum. Wherever it is possible, we don't distinguish between signs of charges for simplicity.

With a frequency $\nu_X$ the charge has a probability of 1/2 to go beyond the cage, to become bared (event (2)). The leaving charge hops over a set of $H_2O$ molecules and through the time $t_U$ is localized, becoming dressed again (event (4)). For the time $\tau_U$ the hydration cage (polarization configuration of water molecules) transfers to the localization distance l. Let the concentration of the bared charges be $N_U$, so that $N_I = N_S + N_U$. The $H_2O$ molecule left by the charge (shown gray in Fig. 3) diffuses as a neutral particle for a time $\tau_W$ (event (5)) until it comes in contact with an ion to be transformed into a hydrated ion (hydration in the scale of $t_W$ happens instantly). The molecular configuration comes back (5') to the initial state (1).

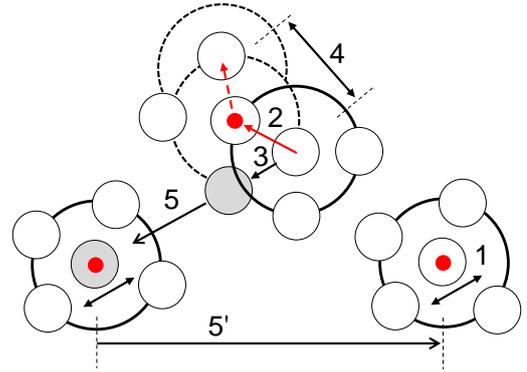

FIG. 4: (Color online) The model of liquid water I-structure. Open circle is an $H_2O$ molecule, red point is an excess proton, point in circle is an $H_3O^+$ ion, shadowed circle is a water molecule left by an excess proton. Big circles are intact (solid) and collapsing (dash) hydrate cages. Arrows show the proton and hydration cage movements. Digits are that used in Fig. 3; they show: (1) - local ion oscillations, (2) - proton hopping, (3) and (5) - diffusion of a neutral water molecule, (4) - hydration cage reconstruction, (5') – ion cloud reconstruction.

The described dynamics makes a continuous process of breaking and forming of $H_3O^+$ and $OH^-$ ions: (2, 3) is collapsing while (4, 5) is in creation. The charges of concentration $N_I$ and neutral $H_2O$ molecules $N_W$ diffuse as separate particles by cycles 2-4-4'-1 ($t_I$) and 3-5-5'-1 ($t_W$), respectively. The part of cycle 1-2 or 1-3, for the time $t_S$, an $H_2O$ molecule and a charge execute an oscillatory motion together being combined into the intact particle (hydrated $H_3O^+$ and $OH^-$ ion) of concentration $N_S$.

In accordance with the above, let us correlate the times $t_W$, $t_U$ and $t_S$ for the dispersion features of the conductivity spectrum $\sigma(\omega)$ in Fig. 2 and 3. The following times can be distinguished as the lifetime of

$t_S = v_S/v_X^2$ – a charge in the intact hydration cage (a dressed charge);

$t_U = 1/v_{D2} - 1/v_X$ – a charge in the collapsing hydration cage (a bared charge);

$t_C \equiv t_I = t_S + t_U$ – a hydration cage (an ion)

$t_W = 1/v_{D1} - 1/v_X$ – a neutral $H_2O$ molecule (in the diffusion mode);

The balance equations for $N_U$, $N_W$ and $N_S$ are:

$$N_W / t_W = N_U / t_U = N_S / t_S \quad (2)$$

where $N_W = N_0 - N_I$ and $N_0 = 55.5$ mol/l $= 3.3 \cdot 10^{28}$ cm$^{-3}$ is a total concentration of particles in liquid water.

The input and calculated parameters of the above scheme are presented in Tables 1 and 2.

TABLE 2: Equilibrium concentration N and the lifetime t of particles in liquid water in accordance with eq. (2) on data specified in Table 1.

|  | Water molecule | Ion bared | Ion dressed | Ion summ |
|---|---|---|---|---|
| Subscript for N | W | U | S | I |
| Subscript for t | W | U | S | C |
| N, $10^{26}$ cm$^{-3}$ | 320 | 7.2 | 7.9 | 15.1 |
| t, $10^{-12}$ s | 49.5 | 1.09 | 1.2 | 2.29 |

Let us next consider the question: what should be the response of the structure shown in Fig. 4 in the conductivity spectrum? We make a natural assumption that diffusion of all particles - neutral molecules, as well as dressed and bared ions obey the law of Brownian diffusion [26]

$$D = \ell^2 / [6(t + t')], \quad (3)$$

where D is a diffusion coefficient, $\ell$ and t are the minimal length and time, only above which the translational steps of a diffusing particle are independent and formula (3) is valid [1]. In other words, $\ell$ and t are Frenkel's elementary steps of diffusion. t' is a delay time caused by departure eq.(3) from normality.

Diffusion of particles in our model is complicated by two accompanying periodic processes (events (4) and (5) in Figs. 3 and 4), namely, by delays of particles at the localized states and by mutual charge transformations. Delays modulate the diffusion rate while transformations modulate the diffusion visibility in the conductivity spectrum.

To simplify the problem, we use the finding made in [2, 3] that in the case of diffusion with periodic stops the residence time t' is added to the time t of normal diffusion. The localization stop-effect decreases the conductivity at frequencies lower $v=1/(t+t')$. Several localization mechanisms of different time scales put a spectrum $\sigma(\omega)$ in the step form. The formula (1) for the $\sigma(\omega)$ spectrum distinguishes the three plateaus, $\sigma_2$, $\sigma_1$, and $\sigma_0$. Let us interpret the plateaus as a response of two spectral steps $\sigma_2 \to \sigma_1$ and $\sigma_1 \to \sigma_0$ as due to the two localization mechanisms – ion hydration and ion ionic screening. The first mechanism is the above mentioned effect of ion dressing with a hydration cage while the second one is additional dressing of a hydrated ion by the cloud of surrounding ions. The second effect is known in physics of electrolytes as an electrophoretic effect [26]. The spectral widths of the steps are defined by the lifetimes of particles in the charged state while the heights are defined by particle mobilities which, in turn, depend on localization times.

We assume that the decay sequence $\sigma_2 \to \sigma_1 \to \sigma_0$ in the direction of frequency decrease $v_X \to v_2 \to v_1$ represents successive dressing of ions with hindering hydration and ion-screening cages. In other words, as frequency decreases a charge reveals itself sequentially as bared, once dressed and twice dressed. The correspondent range of spatial steps is $\lambda \to l \to L$, where $\lambda$ is a step of molecular reconstruction (event (2) in Fig. 4), $l$ is a step of hydration cage reconstruction (event (4)) and L is a step of ionic reconstruction (event (5)). Thus, the dynamics of the ctructure shown in Fig. 4 developes on a hierarchy of space-time scales.

The diffusion of charged particles is observable in the conductivity spectrum in accordance with the Nernst-Einstein's relation [26]

$$\sigma = CND, \quad (4)$$

where $C=q^2/k_BT$, q is an elementary charge, $k_B$ is Boltzmann's Constant, N is the concentration of particles. Frequency dependent conductivity reveals time dependent diffusion. Thus, the plateaus $\sigma_2$, $\sigma_1$, and $\sigma_0$ represent via eq. (4) diffusion of independent bared, dressed and twice dressed particles (the dispersion domains of $\sigma(\omega)$ spectrum reflect movement of particles burdened with reorganization of an environment; under these conditions the formula (4) is unfair). According to the relationships (3) and (4) the diffusion parameters are as follows.

The diffusion coefficient of bared charges is $D_2 = \sigma_2/(CN_U) = 3.1 \cdot 10^{-8}$ m$^2$/c and the diffusion step is $\lambda = 2.99 \cdot 10^{-10}$ m. In this case we assume that only the first step $\lambda$ of a charge at an exit from a cage (point $\nu_X$ in Fig. 3) reflects diffusion of bared charges since immediately after an exit a charge starts acquiring a new cage (puts on).

Similarly, the diffusion coefficient of hydrated (dressed) charges is $D_1 = \sigma_1 / (CN_I) = 8.5$ m$^2$/s and the diffusion step is $l = (6D_1 t_C)^{1/2} = 3.4 \cdot 10^{-10}$ m.

In the case of the dc-plateau $\sigma_0 \equiv \sigma_{dc} = 5.5 \cdot 10^{-6}$ Ohm$^{-1}$m$^{-1}$ a charge is dressed twice, so that the diffusion coefficient is $D_0 = \sigma_0/(CN_I) = 5.9 \cdot 10^{-16}$ m$^2$/s and a characteristic diffusion time is $t_0 = l^2 / (6D_0) = 33 \cdot 10^{-6}$ s.

The values for the sum parameters are shown in Table 3. They connect the model in Fig. 4 via eqs. (1) with the conductivity spectrum in Fig. 2.

TABLE 3: Diffusion parameters in eqs. (3) and (4) of a charge: 2 – purely (partly) bared, 1 – hygrated, 0 – hydrated and additionally ion screened.

| Plateaue number | 0 | 1 | 2 |
|---|---|---|---|
| Subscript | L | 1 | λ |
| $\sigma$, $\Omega^{-1}$m$^{-1}$ | $5.5 \cdot 10^{-6}$ | 80 | 140 |
| D, $10^{-9}$ m$^2$/c | $5.9 \cdot 10^{-7}$ | 8.5 | 31 |
| $\ell$, Å | 13.6 | 3.4 | 2.99 |
| t, ps | $33 \cdot 10^6$ | 50 | 1.1 (1.2) |

## IV. DISCUSSION

1. The dynamics of the structure shown in Fig. 4 is convenient to consider as a process of creation/destruction of $H_3O^+$ and $OH^-$ ions which generates neutral $H_2O$ molecules and charges (protons and holes). The particles thus born migrate for a time. The process goes through thermal collisions. Each ion, having lived ~ 3 ps, transforms into a neutral $H_2O$ molecule, reversely, an $H_2O$ molecule having lived ~ 50 ps transforms into an ion. Accordingly, the $H_2O$ molecules (95.5% of total number of particles) and counterions $H_3O^+$ and $OH^-$ ions (4.5%) are in thermal equilibrium. The ion concentration is close to that obtained by us in [15], and it exceeds by 7 orders of magnitude the common ion concentration, $10^{-7}$% for pH=7.

2. The diffusion step $\lambda = 2.99 \cdot 10^{-10}$ m is, by definition, a distance of molecular structure reconstruction. Given that a charge flows through an $H_2O$ molecule of the diameter d=2.8 Å [7], it moves as combined with the $H_2O$ molecule the distance $\lambda = (2.99^2 - 2.8^2)^{1/2} = 1.0$ Å. This value is exactly a free space distance of the molecular motion in liquid water at $N_W = 55.5$ mol/l.

3. According to the model, the inter-oxygen charge transfer is due to proton hopping. At that time, while the transfer of a *charge* is due to continuous hopping of protons onto a chain, the transfer of a tagged proton (the transfer of *mass*) alternates with stops to wait its turn. Thus, the charge and mass transfers should be distinguished. The former is of a relay-race type, being a few times faster than the latter. The averaged hop probability of a proton from/to $H_3O^+$/$OH^-$ ion is 1/2.4 of that of a charge - thus, the *mass* diffusion coefficient is $D_p = l^2 / [6(t_C + 2.4 \cdot t_C)] = 2.5 \cdot 10^{-9}$ m$^2$/s. It is 3.4 times less than the above *charge* diffusion coefficient $D_2 = 8.5$ m$^2$/s.

4. The distance between the charges of the same sign at $N_I = 1.5 \cdot 10^{27}$ m$^{-3}$ is $L = 2 [ 3 / (2\pi N_I)]^{1/3} = 13.6 \cdot 10^{-10}$ м (from the condition that unit volume is densely packed by the spheres of radius L/2). It is a step of ionic structure reconstruction (event (5') in Fig. 4) wich can be viewed as a step of long-time diffusion both of ions and neutral water molecule. Then, the diffusion coefficient is $D_W = L^2 / [6(t_W + t'_W)] = 2.5 \cdot 10^{-9}$ m$^2$/s where $t'_W = \nu_{D2} \cdot t_C / \nu_{D1}$ is the total delay due to hydration. As is seen, $D_P$ and $D_W$ are equal being both in accordance with data obtained by isotopic tracer method [27, 28].

5. The time found above of 33 μs seems to relate to a category of times, ~ 40 μs, which were obtained by Eigan as a response time of liquid water to pulse impact [29]. Eigan's time is of considerable importance in defining the lifetime of an $H_2O$ molecule in liquid water [30]. The lifetime was established as ~10 hours being assigned to autoionisation of an $H_2O$ molecule. Perception is still undeniably accepted [30, 31]. The point is that this value was obtained under the assumption that the ion concentration is negligibly small, and ions do not interact. In our model the ion concentration is high and the ion-molecular and ion-ion interactions are strong. Under these conditions Eigan's time of 40 μs seems to be due to ionic structure reconstruction rather than an $H_2O$ molecule autoionization. The above model lifetime of an $H_2O$ molecule 50 ps challeng-es the conventional 10 hours, which is a question to be investigated.

6. The spectral 5.3 THz (180 cm$^{-1}$) peak, dominating in Figs. 2 and 3, is commonly known by its ambiguous interpretation [7, 20]. In our model the peak is a straightforward response of $H_3O^+$ or $OH^-$ ion oscillating inside a hydration cage. Given that the ion concentration is high one can conceive that counterions interact to one another forming an ion lattice to be a holder for $H_2O$ molecules. Dynamics of such medium is expected to be due to collective vibrations of protons against frozen lattice rather than to intra-molecular vibrations of separate $H_2O$ molecules. Fast sound could be evidence of the collective acoustic mode [32]. The electrodynamic response should be observable both in the infrared and Raman spectra as is the case [20, 33].

## V. CONCLUSION

We have used an idea that the dielectric spectra of liquid water reflect the translation-oscillator motion of charge particles to construct a gas-solid model of molecular structure of which the dynamics in good agreement with the specra at frequencies $10^3$-$10^{13}$ Hz. According to the model, the dielectric spectra of water are due to diffusion of particles exposed by periodic localizations and recharges. The particles are short-living neutral species $H_2O$ (common $H_2O$ molecules) and charged species $H_3O^+$ and $OH^-$ (ions). The lifetimes of the particles, ~ 50 and ~ 2 ps, respectively, are defined by their mutual transformations during thermal collisions. There are two mechanisms of charge localization - hydration and electrophoretic effect. The mechanism of particle recharge is inter-oxygen proton hops.

The model sheds light on the mobilities of a proton and a neutral water molecule. The problem is that these mobilities are different when recorded in electric experiments dealing with the charge transfer [26] yet equal when recorded in diffusion experiments dealing with the mass transfer [28]. The model clearly shows that charge transfer is of the relay-race type being several times faster than the fair mass transfer. Thus, anomalous proton mobility originating from electric measurements is apparent. It is due to ascribing to one proton of concerted action of several protons.

The characteristic Eigan's time of 40 μs seems to be due to ionic structure reconstruction rather than an $H_2O$ molecule autoionization.

According to the model, a snapshot will show the instantaneous structure of liquid water consisting of molecules $H_2O$ held by counterions of 4.5 % concentration. The Dedye relaxation reflects the motion of ions in the potential of their mutual Coulomb interaction.

High ionic concentration implies an occurrence of the ionic sub-lattice which could explain the puzzling vibrational dynamics of water, - particularly its specific manifestation in infrared and Raman spectra as well as an occurence of fast sound.

By high concentration of ions the model strongly contradicts modern views of water. However, the model is self-consistent and pioneers the link of phenomena which still were separate – the dc conductivity, the Debye and sub-Debye relaxations, as well as the infrared 5.3 THz (180 cm$^{-1}$) peak. It opens wide horizons for further research. If confirmed, the model can be a basis for development of a new paradigm.

\*\*\*